\documentclass[twocolumn, aps,prl,superscriptaddress,showpacs,preprintnumbers,amsmath,amssymb]{revtex4-1}
\usepackage[utf8]{inputenc}
\usepackage{url}
\usepackage{hyperref}
\hypersetup{
    hidelinks = true
}
\usepackage{graphicx,color}
\usepackage{dcolumn}
\usepackage{colordvi} 
\usepackage{bm}
\usepackage{amsmath,amssymb,amsfonts}
\usepackage[]{color}
\usepackage{float}

\begin{document}

\title{LISA pathfinder appreciably constrains collapse models}

\author{Bassam Helou}
\affiliation{Theoretical Astrophysics 350-17, California Institute of Technology, Pasadena, California 91125, USA}

\author{B J. J. Slagmolen}
\affiliation{Australian National University, Canberra, ACT, Australia}
\author{David E.\ McClelland}
\affiliation{Australian National University, Canberra, ACT, Australia}

\author{Yanbei Chen}
\affiliation{Theoretical Astrophysics 350-17, California Institute of Technology, Pasadena, California 91125, USA}
\date{June 2, 2016}

\begin{abstract}
Spontaneous collapse models are phenomological theories formulated to address major difficulties in macroscopic quantum mechanics. We place significant bounds on the parameters of the leading collapse models, the Continuous Spontaneous Localization (CSL) model  and the Diosi-Penrose (DP) model, by using LISA Pathfinder's measurement,  at a record accuracy, of the relative acceleration noise between two free-falling macroscopic test masses.
In particular, we bound the CSL collapse rate to be at most $\left( 2.96 \pm 0.12 \right) \times  10^{-8} \mbox{ s}^{-1}$. This competitive bound explores a new frequency regime, 0.7 mHz to 20 mHz, and overlaps with the lower bound $10^{-8 \pm 2}\mbox{ s}^{-1}$ proposed by Adler in order for the CSL collapse noise to be substantial enough to explain the phenomenology of quantum measurement.
Moreover, we bound the regularization cut-off scale used in the DP model to prevent divergences to be at least $40.1 \pm 0.5 \mbox{ fm}$, which is larger than the size of any nucleus. Thus, we rule out the DP model if the cut-off is the size of a fundamental particle.
\end{abstract}


\maketitle

{\it Introduction.}---
Spontaneous collapse models are modifications of quantum mechanics which have been proposed to explain why macroscopic objects behave classically, and to address the measurement problem. The most widely studied collapse models are the Continuous Spontaneous Localization (CSL) and the Diosi-Penrose (DP) models. 

The CSL model is parametrized by two scales: $\lambda_{{\rm CSL}}$, which sets the strength of the collapse noise, and $r_{{\rm CSL}}$, which sets the correlation length of the noise. For a nucleon in a spatial superposition of two locations separated by a distance much greater than $r_{{\rm CSL}}$, $\lambda_{{\rm CSL}}$ is the average localization rate \cite{AdlerLoUpBounds}. The quantity $r_{{\rm CSL}}$ has  usually been phenomenologically taken to be $100$ nm \cite{BassiReview}, and we will follow this convention. 

The DP model adds stochastic fluctuations to the gravitational field, and is mathematically equivalent to the gravitational field being continuously measured \cite{DiosiConnectionToContinuousMeasTheory1, DiosiConnectionToContinuousMeasTheory2, BassiReview}. The latter statement leaves the DP model with no free parameters, but a regularization parameter, $\sigma_{\rm DP}$, is usually introduced to prevent divergences for point masses. 

Nimmrichter {\it et al.}, in \cite{NimmrichterOMcollapse}, show that the effect of these models on an optomechanical setup, where the center of mass position of a macroscopic object is probed, can be summarized by an additional white noise force, $F\left(t\right)$, acting on the system, and with a correlation function of 
\begin{equation}
\left\langle F\left(t\right)F\left(z\right)\right\rangle =D_{C}\delta\left(t-z\right).
\end{equation} 
For CSL, $D_C$ is given by
\begin{equation}
 D_{\rm CSL} =\lambda_{{\rm CSL}}\left(\frac{\hbar}{r_{{\rm CSL}}}\right)^{2}\alpha
\label{eq:DCSL}
\end{equation}
with $\alpha$ a geometric factor \cite{NimmrichterOMcollapse}. 
LISA pathfinder has quasi-cubic test masses, which we will approximate as perfect cubes. 
For a cube with length $b \gg r_{{\rm CSL}}$,
\begin{equation}
\alpha\approx\frac{8\pi\rho^{2}r_{{\rm CSL}}^4b^{2}}{m_0^2}
\end{equation}
where $\rho$ is the material density, and $m_0$ the mass of a nucleon. For the DP model, $D_C$ is given by
\begin{equation}
D_{\rm DP} \approx\frac{G\hbar}{6\sqrt{\pi}}\left(\frac{a}{\sigma_{{\rm DP}}}\right)^{3} M \rho 
\label{eq:DDP}
\end{equation}
with $M$ the test mass' mass, and $a$ the lattice constant of the material composing the test mass \cite{NimmrichterOMcollapse}.

An optomechanics experiment would need to have very low force noise  to significantly constrain collapse models. LISA pathfinder measures the relative acceleration noise between two free-falling test masses at a record accuracy of $\sqrt{S_a} = 5.2 \pm 0.1 \mbox{ fm s}^{-2}/\sqrt{\rm{Hz}}$ for frequencies between 0.7 mHz and 20 mHz \cite{LISAresults}, and so is a promising platform to test collapse models. We will use $S_{a}$, and relevant details on the LISA pathfinder test mass which we present in table \ref{tab:LPFtestMass}, to provide an upper bound on $\lambda_{{\rm CSL}}$ and a lower bound on $\sigma_{{\rm DP}}$.

We note that $S_a$ has steadily decreased by about a factor of 1.5 since the start of science operations in LISA pathfinder \cite{LISAresults}, and has continued to significantly decrease since the results were published in June 2016 \cite{weberPrivate}. For the remainder of this article, we will use the conservative value of $5.2 \mbox{ fm s}^{-2}/\sqrt{\rm{Hz}}$ for $\sqrt{S_a}$,  but we will also present bounds obtained from a postulated sensitivity level of
$$\sqrt{ S_a^{\rm pos} } = 3.5 \mbox{ fm s}^{-2}/\sqrt{\rm{Hz}},$$
which is about 1.5 times smaller than $\sqrt{S_a}$.

\begin{table}
\caption{\label{tab:LPFtestMass}LISA pathfinder test mass parameters (Ref.
\cite{LISAPathfinderComposition}). We estimated $\rho$
and $a$ with weighted averages of the densities and lattice constants,
respectively, of the materials in the alloy that the test masses are
made out of. The composition of this alloy is 73\% Au and 27\% Pt.}
\begin{ruledtabular}
\begin{tabular}{lcr}
Quantity & Description & Value\tabularnewline
\hline 
$M$ & Mass & 1.928 kg\tabularnewline
$\rho$ & Density & 19881 kg/$\mbox{m}^{3}$\tabularnewline
$a$ & Lattice constant & $4.0$ $\textup{\AA}$\tabularnewline
$b$ & Side length  & 46 mm\tabularnewline
\end{tabular}
\end{ruledtabular}
\end{table}

{\it Constraining the collapse models.}--- We can bound the parameters of collapse models by measuring the force noise of a test mass in an experiment, and attributing unknown noise to the stochastic force $F(t)$. 

In LISA pathfinder, Brownian thermal noise provides the dominant contribution to the differential acceleration noise at frequencies between 1 mHz and 20 mHz. However, the value of this contribution is not precisely known. As a result, we follow a simple and uncontroversial analysis which attributes all acceleration noise to the collapse models' stochastic forces:
\begin{equation}
S_a = 2 S_F / M^2,
\label{SaEq}
\end{equation}
where $S_F = 2 D_C$ is the single sided spectrum of the collapse force. The factor of 2 in Eq. (\ref{SaEq}) follows from the collapse noise on each test mass adding up, because the spontaneous collapse force acts independently on each of the two test masses, which are separated by about 38 cm, a distance much larger than $r_{\rm CSL}$ and $\sigma_{\rm DP}$. Therefore, we can place an upper bound on $D_C$ of: 

\begin{equation}
D_{C} \leq D_{C}^{\rm max} = M^2 S_a / 4.
\end{equation}

Using Eq. (\ref{eq:DCSL}), we can then bound $\lambda_{\rm CSL}$ to
\begin{equation}
\lambda_{{\rm CSL}}\leq \lambda^{\rm max}_{\rm CSL},
\end{equation}
with
\begin{eqnarray}
\lambda_{{\rm CSL}}^{{\rm max}} & = & \frac{m_{0}^{2}}{32\pi\hbar^{2}r_{{\rm CSL}}^{2}}\left(\frac{M}{\rho}\right)^{2}\frac{1}{b^{2}}S_{a}\\
 & = & 2.96\times10^{-8}\;\mbox{s}^{-1},
\end{eqnarray}
where we have substituted in the values shown in table \ref{tab:LPFtestMass} for $\rho$, $M$ and $b$. If we use $S_a^{\rm pos}$ instead of $S_a$, then we reduce $\lambda_{{\rm CSL}}^{{\rm max}}$ to $1.34 \times10^{-8}\;\mbox{s}^{-1}$.

In addition, using Eq. (\ref{eq:DDP}), we can bound $\sigma_{{\rm DP}}$ to
\begin{equation}
\sigma_{{\rm DP}} \geq \sigma_{{\rm DP}}^{\rm min}, 
\end{equation}
with 
\begin{equation}
\sigma_{{\rm DP}}^{{\rm min}}=\left(\frac{2\hbar G}{3\sqrt{\pi}}\frac{\rho}{m}\frac{1}{S_{a}}\right)^{1/3}a=40.1\,\mbox{fm},
\end{equation}
where we have substituted in the values shown in table \ref{tab:LPFtestMass} for $\rho$, $M$ and $a$. If we use $S_a^{\rm pos}$ instead of $S_a$, then we increase $\sigma_{{\rm DP}}^{{\rm min}}$ to $52.2\,\mbox{fm}$.

{\it Discussion.}--- 
LISA pathfinder provides a competitive bound on $\lambda_{\rm CSL}$. $\lambda_{\rm CSL}^{\rm max}$ is three orders of magnitude lower than the bound $10^{-5} \mbox{ s}^{-1}$, which Feldmann and Tumulka \cite{FeldmanDiffBound} calculated from Gerlich {\it et al.}'s  matter wave inteferometry experiment of organic compounds up to 430 atoms large \cite{gerlichInterefomtryExp}. Another matter wave interferometry experiment from the same group \cite{matterInterf2} places a bound of $5 \times 10^{-6} \mbox{ s}^{-1}$, as calculated in \cite{matterInterfCalc2}.

Moreover, $\lambda_{\rm CSL}^{\rm max}$ is comparable to bounds on $\lambda_{\rm CSL}$ obtained from measuring spontaneous heating from the collapse noise. Bilardello {\it et al.} place a bound of $5 \times 10^{-8} \mbox{ s}^{-1}$ \cite{boundFromColdAtomExperiment}, by analyzing the heating rate of a cloud of Rb atoms cooled down to picokelvins \cite{coldAtomExperiment}. Note that Bilardello {\it et al.}'s bound depends on the temperature of the CSL noise field, and on the reference frame with respect to which the CSL noise field is at rest with \cite{boundFromColdAtomExperiment}. The standard formulation of CSL has the collapse noise field at a temperature of infinity, but the theory could be modified to include different temperatures. The incorporation of dissipation within CSL is based on the dissipative CSL (dCSL) theory proposed by Smirne and Bassi \cite{dCSL}.

Other competitive upper bounds have been obtained from cosmological data, the lowest of which, $10^{-9} \mbox{ s}^{-1}$, is from the heating of the intergalactic medium \cite{AdlerLoUpBounds}. However, this bound is also sensitive to the temperature of the collapse noise field \cite{dCSL}. Moreover, our interest in this article is for controlled experiments.

\begin{figure}
\centering\includegraphics[scale=0.47]{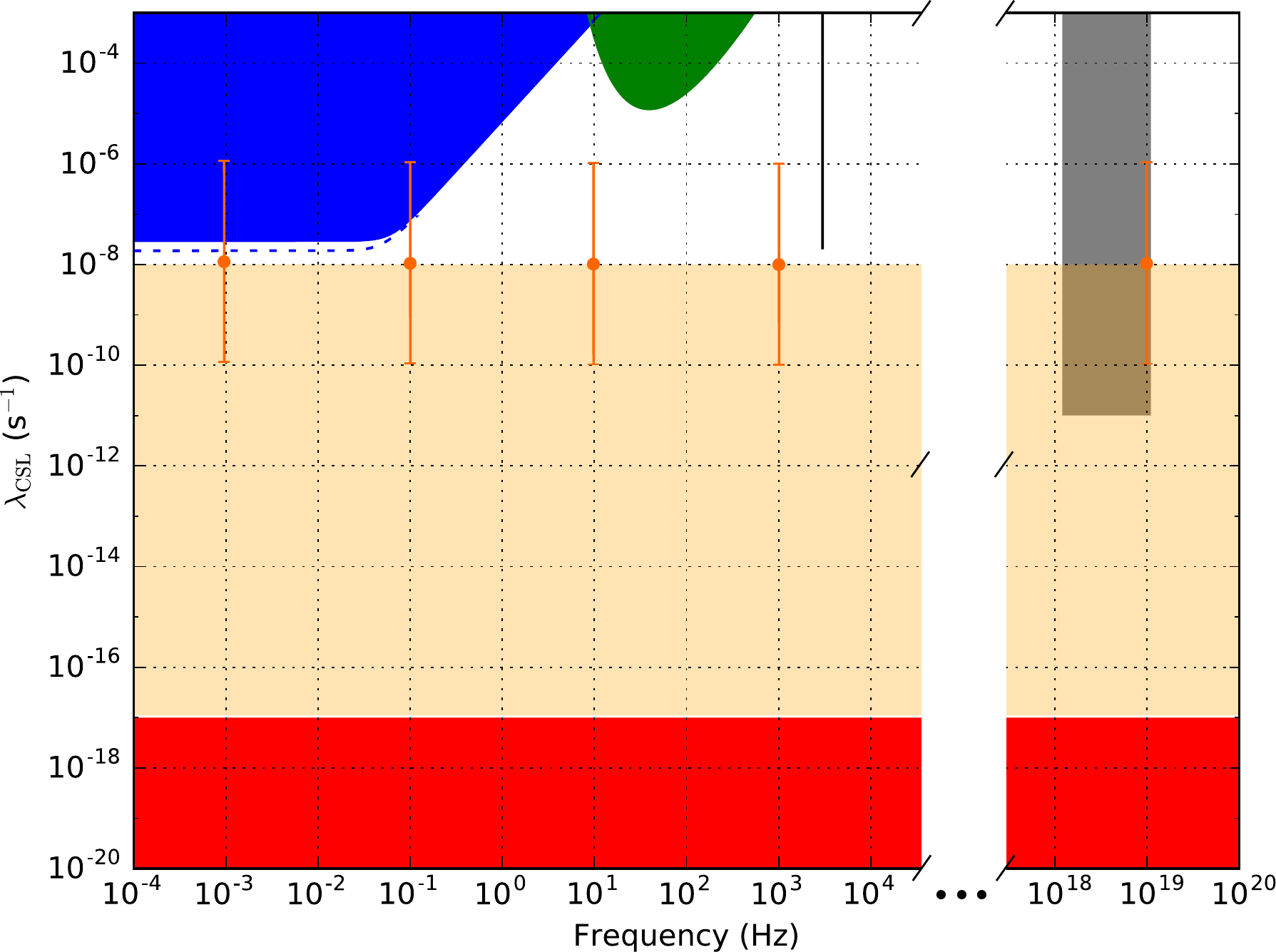}

\caption{\label{fig:exclusionPlot}Upper and lower bounds on the CSL collapse rate $\lambda_{\rm CSL}$ obtained from laboratory experiments operating at different frequencies. Blue, green, black and gray regions: exclusion regions obtained from LISA pathfinder, LIGO, a millikelvin-cooled nanocantilever \cite{upperBoundFromCantilever} and spontaneous emission from Ge \cite{limitXray, xrayNew}, respectively. Our calculation of the bounds obtained from LIGO follow that of \cite{competitorLISApaper}. The dashed blue line is the upper bound limit obtained from the LISA pathfinder results if $S_a^{\rm pos}$ were used instead of $S_a$.
The red and orange domains are regions in which the collapse rate is too slow to explain the lack of macroscopic superpositions and measurements, respectively.
The red region is below the lower bound  of $10^{-17}\mbox{ s}^{-1}$ proposed by Ghirardi, Pearle and Rimini \cite{GhiradiLowerBound}.
The orange region's boundary is the Adler lower bound  $10^{-8 \pm 2}\mbox{ s}^{-1}$, below which latent image formation on a photographic emulsion consisting of silver halide suspended in gelatine wouldn't occur  fast enough \cite{AdlerLoUpBounds}. The orange error bars reflect the uncertainty in this lower bound. }
\end{figure}

In addition to providing an aggressive upper bound, LISA pathfinder explores the low frequency regime of 0.7 mHz to 20 mHz. In Fig. \ref{fig:exclusionPlot}, we compare $\lambda_{\rm CSL}^{\rm max}$ to bounds obtained from experiments operating in different frequency regimes. If $S_a^{\rm pos}$ is used instead of $S_a$, then LISA pathfinder provides the smallest upper bound of all experiments operating below a THz scale.

LIGO's measurement of the differential displacement noise between two test masses in the frequency regime 10 Hz to 10 kHz places upper bounds of at most about $10^{-5} \mbox{ s}^{-1}$. In \cite{upperBoundFromCantilever}, an upper bound of about $2\times 10^{-8} \mbox{ s}^{-1}$ is obtained by analyzing the excess heating of a nanocantilever's fundamental mode at about 3.1 kHz. 
A record upper bound of $10^{-11} \mbox{ s}^{-1}$ is placed in \cite{limitXray, xrayNew} by examining the spontaneous x-ray emission rate from Ge. This bound could be greatly reduced if the collapse noise is non-white at the very high frequency of $10^{18} \mbox{ s}^{-1}$ \cite{BassiReview}. 

Furthermore, the bound $\lambda_{\rm CSL}^{\rm max}$ appreciably constrains the CSL model because it overlaps with some of the proposed lower bounds on $\lambda_{\rm CSL}$.
Adler investigates the measurement process of latent image formation in photography and places a lower bound of $\lambda_{\rm CSL} \simeq 2.2 \times 10^{-8\pm2} \mbox{ s}^{-1}$ \cite{AdlerLoUpBounds}. Moreover, Bassi {\it et al.} place a lower bound of $\lambda_{\rm CSL} \simeq 10^{-10\pm2} \mbox{ s}^{-1}$ by investigating the measurement-like process of human vision of six photons in a superposition state \cite{BassiEye}. Note that a lower bound of about $10^{-17} \mbox{ s}^{-1}$, proposed by Ghirardi, Pearle and Rimini \cite{GhiradiLowerBound}, is also sometimes considered. Its justification comes from the requirement that an apparatus composed of about $10^{15}$ nucleons settle to a definite outcome in about $10^{-7} \mbox{ s}$ or less \cite{AdlerScience}.

LISA pathfinder also provides a competitive bound on $\sigma_{{\rm DP}}$. 
The nanocantilever experiment \cite{upperBoundFromCantilever} places a lower bound on $\sigma_{\rm DP}$ of about 1.5 fm, which is much lower than $\sigma_{{\rm DP}}^{\rm min}$. 
More importantly, the calculated value for $\sigma_{{\rm DP}}^{\rm min}$ of $40.1 \pm 0.5 \mbox{ fm}$ is larger than the size of any nucleus. Consequently, $\sigma_{{\rm DP}}^{\rm min}$ rules out the DP model if the regularization scale $\sigma_{\rm DP}$ is chosen to be the size of a fundamental particle such as a nucleon.


{\it Acknowledgments.}---  We acknowledge support from the NSF grants PHY-1404569 and PHY-1506453, from the Australian Research Council grants FT130100329 and DP160100760, and from the Institute for Quantum Information and Matter, a Physics Frontier Center.


\bibliography{references}

\end{document}